\documentclass{aa}

\usepackage{graphicx}

\usepackage{natbib}
\usepackage{txfonts}

\begin{document}

\title{The origin of abundance gradients in the Milky Way: the predictions of different models} 

\titlerunning{Chemical evolution along radius}

\author {E. Colavitti\inst{1,}\thanks{email to: colavitti@oats.inaf.it}
\and G. Cescutti\inst{1}
\and  F. Matteucci\inst{1,2}
\and  G. Murante\inst{3}}

\authorrunning{E. Colavitti et al.}

\institute{Dipartimento di Astronomia, Universit\'a di Trieste,  Via G. B. Tiepolo 11, I-34143 Trieste (TS), Italy 
\and  I.N.A.F. Osservatorio Astronomico di Trieste, Via G. B. Tiepolo 11, I-34143 Trieste (TS), Italy
\and  I.N.A.F. Osservatorio Astronomico di Torino, Strada Osservatorio 20, I-10025 Pino Torinese (TO), Italy}

\date{Received xxxx / Accepted xxxx}

\abstract{}
{We aim at studying the abundance gradients along the Galactic disk and their dependence upon several parameters: a threshold in the surface gas density regulating star formation, the star formation efficiency, the timescale for the formation of the thin disk and the total surface mass density of the stellar halo. We use chemical evolution models already tested on a large number of observational constraints.} 
{We test a model which considers a cosmological infall law. This law does not predict an inside-out disk formation, but it allows to well fit the properties of the solar vicinity. To check whether this model can reproduce the properties of the galactic disk, we study several cases. We drop the threshold in the surface gas density and assume a star formation efficiency varying with radius. We also test the same parameters in the two-infall model for the Galaxy. Finally, we perform some additional analysis on our simulations to test if the cosmological infall law can account for the inside-out formation of the disk.}  
{We find that to reproduce at the same time the abundance, star formation rate and surface gas density gradients along the Galactic disk it is necessary to assume an inside-out formation for the disk. The threshold in the gas density is not necessary and the same effect could be reached by assuming a variable star formation efficiency. The derived new cosmological infall law contains a mild inside-out formation and is still not enough to reproduce the disk properties at best.}
{A cosmologically derived infall law with an inside-out process for the disk formation and a variable star formation efficiency can indeed well reproduce all the properties of the disk. However, the cosmological model presented here does not have sufficient resolution to capture the requested inside-out formation for the disk. High resolution cosmological simulations should be performed to better capture such a behavior.} 

{}

\keywords{Galaxy: evolution, Galaxy: formation, Galaxy: disk, Galaxy: abundances}

\maketitle

\section{Introduction}

Understanding the formation and the evolution of  the Milky Way is fundamental to improve the knowledge of the formation of spiral galaxies in general. In many models of the chemical evolution of the Milky Way gas infall has been invoked to explain the formation of a galactic disk (e.g. Chiosi 1980; Lacey \& Fall 1985; Matteucci \& Fran\c cois 1989; Chiappini et al. 1997 (hereafter C97); Boissier \& Prantzos 2000, Portinari \& Chiosi 2000 among others). Originally, the gas infall was introduced as a possible solution to the G-dwarf problem (Pagel 1989). The infall law is clearly very important to determine the main characteristics of a galaxy. 

Our Galaxy has many observational constraints. Some of the most important ones are represented by the evolution of the abundances of the chemical elements. Recently, many chemical evolution models have been developed to explain the chemical composition of the solar vicinity (e.g. Henry et al. 2000; Liang et al. 2001; Chiappini et al. 2003a, 2003b; Akerman et al. 2004; Fran\c cois et al. 2004; Cescutti et al. 2006, 2007). Other important constraints are the abundance gradients of the elements along the disk of the Milky Way. Abundance gradients are observed in many galaxies with their metallicities decreasing outward from the galactic centers.  They are strongly influenced by the star formation and by the accretion history as functions of galactocentric distance (see Matteucci \& Fran\c cois 1989; Boissier \& Prantzos 1999; Chiappini et al. 2001; Colavitti et al. 2008). In particular, Colavitti et al. (2008) (hereafter C08) adopted an infall law  cosmologically motivated, in the sense that assumed an accretion law for baryons which is similar to the accretion law of dark matter, as derived by cosmological simulations, obtained  with the public tree-code GADGET2 (Springel 2005). This cosmological law is able to well reproduce all the constraints in the solar vicinity, such as the star formation rate, the supernova (SN) Ia and SNII rate, the total amount of gas and stars and the behavior of several chemical elements (such as the $\alpha$-elements, C, N and Fe).

In this paper we base our work on two chemical evolution models: the one of C97 and the one of C08. The C08 model does not predict an inside-out formation of the thin disk as in the C97 model and while the C08 model predicts good star formation rate and gas density distributions as functions of the galactocentric distance, it does predict a too flat slope for the oxygen gradient in the inner disk regions.

The aim of this paper is to test the predictions of the C08 model along the disk of the Milky Way by varying a number of important parameters in order to understand if there are alternative solutions to the inside-out formation and the threshold in the gas density for star formation to reproduce abundance gradients. In particular, we test several cases in which we vary: i) the threshold gas density during the formation of the halo, thick and thin disk components, ii) the efficiency of star formation as a function of galactocentric distance, iii) the total surface mass density of the stellar halo and iv) the timescale for the formation of the thin disk. 
We do not test the effects of a variable initial mass function (IMF) along the disk, since it has already been done, and it has been concluded that it cannot reproduce at the same time the abundance gradients, the gas and star formation rate distributions (e.g. Carigi 1996; Chiappini et al. 2000). Radial flows can also play a role in the formation of abundance gradients (see Portinari \& Chiosi 200, and references therein). In particular, radial flows can produce gradients but mainly in the outer regions of the disk while the inner gradients are too flat.
This effect is similar to that of the threshold in the gas density without an inside-out formation and therefore we do not take it into account here. What is more important is the effect produced by a bar in the bulge which can explain the peak of the gas observed at 4-5 Kpc from the Galactic center (Portinari \& Chiosi 2000), contrarily to models which do not include the bar. However, we are not interested in this problem here but simply to the gradients and gas and star formation rate distributions for a galactocentric distance $R_G> 4$ kpc.

We compare our model predictions to observational data relative to abundance gradients along the disk: those of Andrievsky et al. (2002a-c, 2004) and Luck et al. (2003). They measured the abundances of all the selected elements (O, N and Fe) in a sample of 130 Galactic Cepheids found in the galactocentric distance range from 5 to 17 kpc. Moreover, we compare our results with the observational data collected by Simpson et al. (1995), Afflerbach et al. (1997) and Gummersbach et al. (1998) for the N gradient from HII regions and with those collected by Chen, Hou \& Wang (2003), Twarog et al. (1997) and Carraro et al. (1998) for the Fe gradient from open clusters. We also compare our model results with the star formation rate and the surface gas density as functions of the galactocentric distance by adopting the data collected by Rana (1991). He took the average $HI$ distribution from Blitz et al. (1983), Gordon \& Burton (1976) and Garwood \& Dickey 1989 and two distinct $H_{2}$ distributions: one from Bronfman (1986) and the other from Robinson et al. (1988), Wouterloot et al. (1990) and Digel et al. (1990). Moreover he took the radial dependences of the present surface density of the star formation rate from Rana \& Wilkinson (1986). 

The paper is organized as follows: in section 2 we show the observational data and nucleosynthesis prescriptions adopted. Section 3 presents a brief description of the model by C97. In section 4 we describe the C08 model and the cosmological simulations. Section 5 describes the results obtained, comparing the models predictions with the observed properties. Finally section 6 presents the conclusions.  

\section{Observational data and nucleosynthesis prescriptions}

In this work we use the data by Luck et al. (2003) and  Andrievsky et al. (2002a-c, 2004) for all the studied elemental gradients. These data have been derived for a large sample of Galactic Cepheids. These variable stars have a distinct role in the determination of radial abundance gradients for several reasons. First, they are usually bright enough that they can be observed at large distances, providing accurate abundances; second, their distances are generally well determined, as these objects are often used as distance calibrators; third, their ages are also well determined. They generally have ages close to a few hundred million years. So we can assume that they are representative of the present day gradients.

Here we also use the data by Gummersbach et al. (1998) (B stars), Simpson et al. (1995) and Afflerbach et al. (1997) from HII regions, and by Chen, Hou \& Wang (2003), Carraro et al. (1998) and Twarog et al. (1997) from open clusters. 

One of the most important ingredients for chemical evolution models is represented by the nucleosynthesis prescriptions and, consequently, by the stellar yields. 

The single stars in the mass range $0.8 \; M_{\odot} \; \leq \; M \; \leq \; 8 \; M_{\odot}$ (low and intermediate-mass stars) contribute to the Galactic enrichment through planetary nebula ejection and quiescent mass loss. They enrich the interstellar medium mainly in He, C,  N and heavy s-process elements (e.g. Cescutti et al. 2006). 
We adopt here the stellar yields for low and intermediate mass stars of van den Hoek \& Groenewegen (1997) computed as functions of stellar metallicity, their case with variable mass loss. Concerning N production, the yields of van den Hoek \& Groenewegen (1997) from AGB stars predict both primary and secondary N. For N produced in massive stars we assume it secondary, although there is some indication that massive stars might produce primary nitrogen (e.g. Meynet \& Maeder 2002)
A small fraction of these stars are also the progenitors of Type Ia SNe if they are in binary systems: they originate from carbon deflagration of C-O white dwarfs. We adopt in this paper the single-degenerate progenitor scenario (Whelan \& Iben, 1973; Han \& Podsiadlowski 2004). Type Ia SNe contribute a substantial amount of Fe ($\sim 0.6 \; M_{\odot}$ per event) and Fe-peak elements as well as non negligible quantities of Si and S. They also produce other elements, such as O, C, Ne, Ca, Mg and Ni, but in very small amounts compared to Type II SNe. We assume the stellar yields for Type Ia SNe from Iwamoto et al. (1999).

Massive stars ($8 \; M_{\odot} \; < \; M \; \leq \; 100 \; M_{\odot}$) are the progenitor of  core-collapse SNe which can be either Type II SNe or Type Ib/c SNe. These latter can arise either from binary systems or Wolf-Rayet stars, whereas Type II SNe originate from the massive stars in the lower mass range (Calura \& Matteucci 2006). Type II SNe mainly produce the so called $\alpha$-elements, such as O, Mg, Ne, Ca, S, Si and Ti, but also some Fe and Fe-peak elements although in smaller amounts than Type Ia SNe.  We adopt here the stellar yields for massive stars by Woosley \& Weaver (1995) with the suggested modifications of Fran\c cois et al. (2004). 
However, the most important modifications concern some Fe-peak elements, except Fe itself, whereas for the $\alpha$-elements, with the exception of Mg which has been increased relative to the original yields, the yields are substantially unmodified relative to Woosley \& Weaver's ones.

The reference solar abundances are those by Asplund et al. (2005).       

\section{The model by Chiappini et al. (1997)}

The model by C97 was the first in which two main infall episodes for the formation of the Galactic components were suggested. In particular, they assumed that the first infall episode was responsible for the formation of the halo and thick-disk stars that originated from a fast dissipative collapse. The second infall episode formed the thin-disk component, with a timescale much longer than that of the thick-disk formation. The timescale for the formation of the halo-thick disk is 0.8 Gyr, while the timescale for the thin disk is 7 Gyr in the solar vicinity. The authors included in the model also a threshold in the gas density, below which the star formation process stops. The existence of such a threshold value is suggested by observations relative to the star formation in external disk galaxies (Kennicutt 1998, but see Boissier et al. 2006). The physical reason for a threshold in the star formation is related to the gravitational stability, according to which, below a critical density, the gas is stable against density condensations and, consequently, the star formation is suppressed. In the two-infall model the halo-thick disk and the thin disk evolutions occur at different rates and they are independent, mostly as a result of different accretion rates. With these precise prescriptions it is possible to reproduce the majority of the observed properties of the Milky Way and this shows how important is the choice of the accretion law for the gas coupled with the star formation rate in the Galaxy evolution. Some of the most important observational constraints are represented by the various relations between the abundances of metals (C, N, $\alpha$-elements, iron peak elements) as functions of the [Fe/H] abundance and by the G-dwarf metallicity distribution. In C97 the Galactic disk is approximated by a series of concentric annuli, 2 kpc wide, without exchange of matter between them. In this model the thin disk forms ``inside-out'', in the sense that it assumes that the timescale for the disk formation increases with galactocentric distance. This choice was dictated by the necessity of reproducing the abundance gradients along the Galactic disk. The SFR is a Schmidt (1955) law with a dependence on the surface gas density ($k=1.5$, see Kennicutt 1998) and also on the total surface mass density (see Dopita \& Ryder 1994). In particular, the SFR is based on the law originally suggested by Talbot \& Arnett (1975) and then adopted by Chiosi (1980):

\begin{equation}
\psi(r,t)=\nu\left(\frac{\Sigma(r,t) \Sigma_{gas}(r,t)}{\Sigma(r_{\odot},t)^{2}}\right)^{(k-1)}\Sigma_{gas}(r,t)^{k}
\end{equation}

where $t$ is the time and the constant $\nu$ is a sort of efficiency of the star formation process and is expressed in $Gyr^{-1}$: in particular,  $\nu= 2 \,Gyr^{-1}$ for the halo and $1 \, Gyr^{-1}$ for the disk ($t\ge 1\, Gyr$). The total surface mass density is represented by $\Sigma(r,t)$, whereas $\Sigma(r_{\odot},t)$ is the total surface mass density at the solar position, assumed to be $r_{\odot}=8$ kpc (Reid 1993). The quantity $\Sigma_{gas}(r,t)$ represents the surface gas density and $t$ represents the time. These choices of values for the parameters allow the model to fit very well the observational constraints, in particular in the solar vicinity. A threshold gas density for the star formation in the thin disk of $7 M_{\odot} pc^{-2}$ is adopted in all the models presented here. The IMF is that of Scalo (1986) normalized over a mass range of 0.1-100 $M_{\odot}$ and it is assumed to be constant in space and time. In this paper we will adopt also a simple Schmidt law where the SFR depends only on the surface gas density:

\begin{equation}
\psi(r,t)=\nu \; \Sigma_{gas}(r,t)^{k}
\end{equation} 

where k = 1.5 and $\nu$ is proportional to 1/R.

\section{The model of Colavitti et al. (2008)}

The model of C08 is basically the one of C97 but with an accretion law derived from cosmological simulations.
In particular, we run a dark matter-only cosmological simulation, using the public tree-code GADGET2 (Springel 2005), in order to produce and study dark matter halos in which spiral galaxies can form. Our simulated box has a side of 24 $h^{-1}$ Mpc. We used $256^{3}$ particles.  We adopted the standard cosmological parameters from WMAP 3-years (Spergel et al. 2007), namely $\Omega_{0} = 0.275$, $\Omega_{\lambda} = 0.725$ and $\Omega_{b} = 0.041$. Each DM particle has a mass equal to $6.289 \cdot 10^{7} h^{-1} M_{\odot}$ and the Plummer-equivalent softening length is set to 3.75 $h^{-1}$ comoving kpc to redshift $z=2$ and to $1.25 \,h^{-1}$ physical kpc since $z=2$. We use the public package GRAFIC (Bertschinger 1995) to set up our initial conditions. The simulation started at redshift $z=20$ and 28 outputs were produced. We have chosen to use a quite large spread in the redshifts at the beginning, while in the last part of the simulation, where a small change in the redshift corresponds to a large change in time, the redshifts are closer. We checked that the final mass function of DM halos and the power spectrum are in agreement with theoretical expectations.

We identified DM halos at redshift $z=0$ using a standard friend-of-friends algorithm, with a linking length $l=0.17$ mean (comoving) interparticle distance. After that, we determined the virial mass and radius for each DM halo, using the center of mass of the F-o-F group as the halo center. Here we define the virial radius as the radius of the sphere within which the matter density contrast is $\delta \approx 100$ times the critical density, with $\delta$ given by the cosmological parameter as in Navarro \& Steinmetz (2000).

We then built the mass accretion history of our halos. To achieve this goal, we analysed the 28 outputs obtained from redshift $z = 9.0$ to $z = 0$. We identified all DM halos in each snapshot using the procedure described above, except for the fact that we used the redshift-dependent density contrast given by Bryan \& Norman (1997) to define the virial radius as a function of $z$. At any output $z_{i+1}$, we found all the progenitors of our halos at redshift $z_i$. We defined a halo at redshift $z_{i+1}$ to be a progenitor of one at $z_i$ if at least 50\% of its particles belong to the candidate offspring (see e.g. Kauffmann 2001, Springel et al. 2001 for a discussion of this threshold). The mass accretion history is defined as the mass of the main progenitor of the halo as a function of redshift.  Having the mass accretion histories, we were able to identify the redshift of formation (defined as the epoch at which half of the mass of the forming halos were accreted) and the redshift at which each halo experienced its last major merger (defined as an increase of at least 25\% of its mass with respect to the mass of its main progenitor at the previous redshift). To identify the DM halos that can host a spiral galaxy similar to the MW we used selection criteria based on four different characteristics of the halos:

\begin{itemize}
 
\item mass between $5 \cdot 10^{11} M_{\odot}$ and $5 \cdot 10^{12} M_{\odot}$;

\item spin parameter $\lambda > 0.04$; 

\item redshift of last major merger ($z_{mm}$) larger than $z=2.5$;

\item redshift of formation ($z_{f}$) larger than $z=1.0$.

\end{itemize}

In particular, in this paper we use our best cosmological halo which has these characteristics: 

\begin{itemize}

\item M = $90.26 \cdot 10^{10} M_{\odot}$; 

\item $\lambda$ = 0.045; 

\item $z_{mm}$= 5.00; 

\item 1.50 $<$ $z_{f}$ $<$ 1.75. 

\end{itemize}

Assuming that the baryonic matter follows the same accretion pattern as the dark matter, and that it represents 19\% (the cosmological baryon fraction) of all the infalling matter, we obtained a final baryonic mass for the Galaxy of $1.7 \cdot 10^{11} M_{\odot}$. We assumed that the derived infall law has the same functional form for the whole Milky Way, but that the normalization constant is different for different Galactic regions. In other words, the normalization constants were obtained by reproducing  the current total surface mass density at any specific galactocentric distance.

\begin{figure}
\centering
\includegraphics[width=0.45\textwidth]{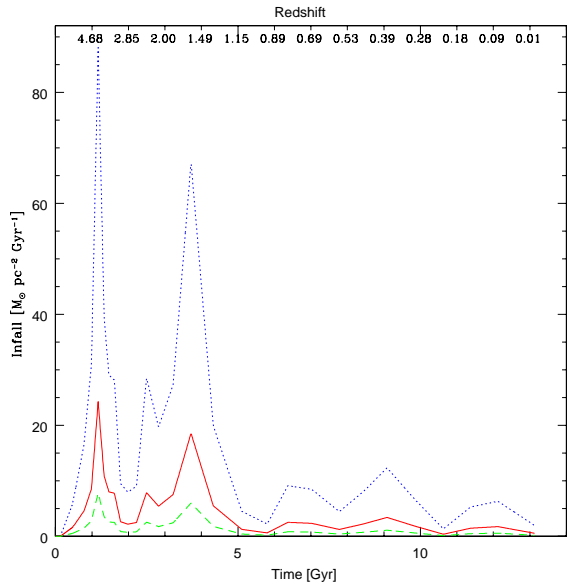}
\caption[]{Cosmological infall laws at 4 (blue dotted line), 8 (red solid line) and 14 kpc (green dashed line).}
\label{infall}
\end{figure}

\section{Results}

In this work we tested several different models in order to understand if it is possible to well reproduce all the observed values just acting on four variables. In particular, in C08 (Model C08) we saw that our best cosmological model was not able to reproduce the [O/H] ratio in the inner Galactic disk. So in this paper we try to find the better combination of variables to reproduce the oxygen gradient. We also test these variables in the model by C97. These are the considered variables:

\begin{itemize}

\item \textbf{Threshold.} In the original model by C97 the threshold in the surface gas density was equal to 4 $M_{\odot} \; pc^{-2}$ during the formation of the halo and to 7 $M_{\odot} \; pc^{-2}$ during the formation of the thin disk. When the surface gas density was lower than these values the SFR stopped. In this paper we switch off the threshold in all the models.

\item \textbf{SF efficiency ($\nu$).} In C97 and in C08 the star formation efficiency was assumed to be constant along the disk. Here we use, for some models, a SF efficiency which is a function of radius. This approach is already known (see Boissier \& Prantzos 1999) and such a dependence has been proposed on the basis of large-scale instabilities in rotating discs. Even in Carigi (1996) the author used a model in which the star formation efficiency was variable with radius ($\nu(r) \propto r^{-1}$, as already seen in Prantzos \& Aubert 1995). We adopt several ad hoc variations of the parameter $\nu$ as a function of radius plus the law adopted in Boissier \& Prantzos (1999). All the adopted laws are shown in Table 1.

\item \textbf{Timescale for the formation of the disk ($\tau$).} In C97 the Galactic disk forms inside-out. This means that the inner parts of the galactic disk formed first than the outer ones. This type of formation was introduced to explain the Galactic gradients. Even in Carigi (1996) the author used a model in which the infall timescale for the disk was variable with radius. The model by C08, instead, did not use an inside-out scenario since the cosmologically derived infall law does not contain such process. So in this case the parameter $\tau$ cannot be changed. In Figure 1 we show the cosmological infall law of C08 at different radii. Clearly the amount of gas falling at each radii is different, but the shape of the law is maintained. 

In some of the models we use a constant timescale while in others we use the prescription used in C97, namely:

\begin{equation}
\tau = 1.03 \; r - 1.27 \; [Gyr]. 
\end{equation}  

where $r$ is the galactocentric radius in kpc. Other authors have also studied an inside-out formation of the disk by assuming a linear dependence of $\tau$ on $r$ (e.g. Matteucci \& Fran\c cois 1989, Carigi 1996, Boissier \& Prantzos 1999).

\item \textbf{Surface halo mass density ($\Sigma_{halo}$).} In C97 the projected halo mass density was a constant parameter. Here we use both a halo density which is a function of radius and a constant one, as in Chiappini et al. (2001). In particular, it follows this equation:

\begin{equation}
\Sigma_{halo} = \frac{136}{r} \; [M_{\odot} \; pc^{-2}]. 
\end{equation} 

where $r$ is the radius in kpc. The value of 136 is adopted in order to have a surface halo mass density equal to 17 $M_{\odot} \; pc^{-2}$ at the solar neighbourhood. The halo density is important in determining the slope of the gradient at large galactocentric distances, where the disk density is very low (see Chiappini et al. 2001).

\end{itemize}     

In Table 1 we present the models and all the parameters which characterize them. The second column indicates the presence or the absence of a threshold, the third column the type of variation of the SF efficiency (the values of $\nu$ for Model 3 are for 2.5, 7.5 and 12.5 kpc), the fourth shows if the galaxy forms inside-out or not and the fifth if the halo density is constant or if it is a function of radius. In the sixth column we show the type of infall law used and in the seventh column the SFR. 

\begin{table*}
\caption{Models parameters. The second column indicates the presence or the absence of a threshold, the third column the type of SF efficiency, the fourth if the galaxy forms inside-out and the fifth if the halo density is constant or if it is a function of radius. In the sixth column we show the type of infall law used and in the seventh column the SFR. The values of $\nu$ for Model 3 are for 2.5, 7.5 and 12.5 kpc.}
\centering
\begin{tabular}{c|c|c|c|c|c|c}
\noalign{\smallskip}
\hline
\hline
\noalign{\smallskip}
Model & Threshold & $\nu$ (4 - 6 - 8 - 10 - 12 - 14 kpc) & $\tau$ & $\Sigma_{halo}$ & Infall law & SFR \\
\noalign{\smallskip}
\hline
\noalign{\smallskip}
C08 & yes & const & / & / & cosmological & eq. (1) \\
C97 & yes & const & 1.03 $\cdot$ R - 1.27 & const & two-infall law & eq. (1) \\
1 & no & const & / & / & cosmological & eq. (1) \\
2 & no & 8.0 - 4.0 - 1.0 - 0.5 - 0.2 - 0.05 & / & / & cosmological & eq. (1) \\
3 & no & 4.0 - 1.0 - 0.1 & / & / & cosmological with inside-out & eq. (1) \\
4 & no & const & const & const & two-infall law & eq. (1) \\
5 & no & const & 1.03 $\cdot$ R - 1.27 & const & two-infall law & eq. (1) \\
6 & no & 9.0 - 3.0 - 1.0 - 0.3 - 0.1 - 0.03 & 1.03 $\cdot$ R - 1.27 & $\propto$ 1/R & two-infall law & eq. (1) \\
7 & no & $0.1 \cdot 8/R$ & const & / & one-infall law & eq. (2) \\
\noalign{\smallskip}
\hline
\hline
\end{tabular}
\end{table*} 

Model C08 is the best cosmological model of Colavitti et al. (2008). It had the threshold only during the formation of the disk and a constant star formation efficiency $\nu$. As explained above, with this model the authors were not able to reproduce the observed values for the [O/H] in the inner part of the disk. 

Model C97 is the original model of C97, i.e. the two-infall model. It has a threshold, both during the formation of the halo and thick-disk component and during the formation of the thin disk. It has a star formation efficiency constant along radius and equal to $\nu = 1.0 \; Gyr^{-1}$, an inside-out prescription and a constant surface halo density along radius.

\subsection{Results for models C08, C97, 1 and 2}
\begin{figure}
\centering
\includegraphics[width=0.45\textwidth]{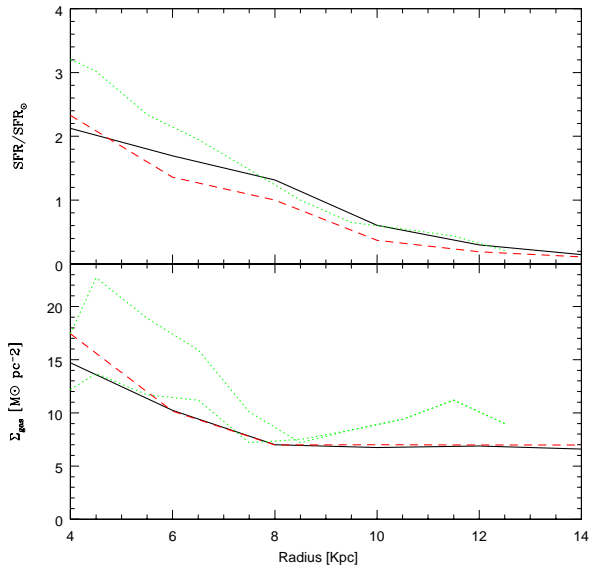}
\caption[]{$SFR/SFR_{\odot}$ and surface gas density vs radius for Model C97 (black solid line) and for Model C08 (red dashed line). The data are from Rana (1991) (green dotted lines, representing the lower and the maximum values for the gas).}
\label{C97.1}
\end{figure}

\begin{figure}
\centering
\includegraphics[width=0.45\textwidth]{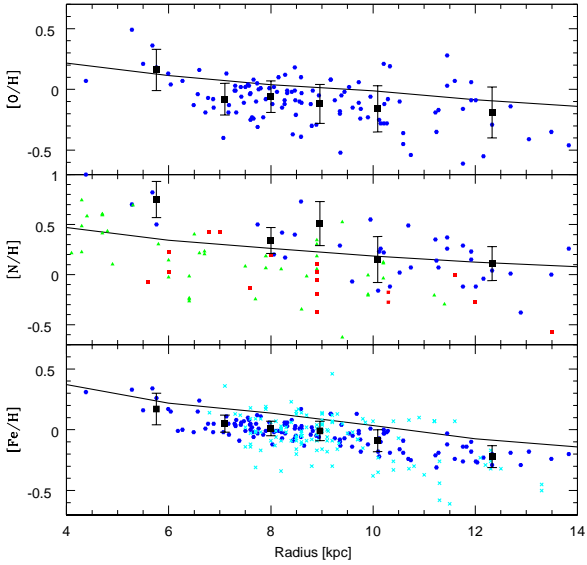}
\caption[]{[O/H], [N/H] and [Fe/H] vs radius for Model C97 (black solid line) and for Model C08 (red dashed line). The data are from a compilation by Cescutti et al. (2007) (blue dots, Cepheids), from Gummersbach et al. (1998) (red squares, B stars), from Simpson et al. (1995) and Afflerbach et al. (1997) (green triangles, HII regions) and from Chen, Hou \& Wang (2003), Carraro et al. (1998) and Twarog et al. (1997) (cyan crosses, open clusters). The black squares are the mean values inside each bin only for the compilation by Cescutti et al. (2007) and the error bars are the standard deviations.}
\label{C97.2}
\end{figure}

In Figures \ref{C97.1} and \ref{C97.2} the results for Model C97, compared with those predicted by Model C08 and with observations, can be seen. Figure \ref{C97.1}, upper panel, presents the SFR normalized at the solar neighbourhood value. The data are from Rana (1991) (green dotted line) while the red dashed line represents Model C08. The black solid line represents Model C97. It can be seen that both Model C97 and Model C08 have no SF in the outer part of the disk, both having a threshold in the surface gas density below which the SF stops. The bottom panel of Figure \ref{C97.1} shows the current surface gas density along the radius. The data are, once again, from Rana (1991). In this case Model C97 and Model C08 produce very similar results. Moreover both the models reproduce very well the lower values of Rana (1991). 

Figure \ref{C97.2} shows the evolution of [O/H], [N/H] and [Fe/H] along the radius for Models C97 and C08. The model results are normalized to the Asplund et al. (2005) solar abundances in all the models. The data are from a compilation by Cescutti et al. (2007) (blue dots, Cepheids stars), from Gummersbach et al. (1998) (red squares, B stars), from Simpson et al. (1995) and Afflerbach et al. (1997) (green triangles, HII regions) and from Chen, Hou \& Wang (2003), Carraro et al. (1998) and Twarog et al. (1997) (cyan crosses, open clusters). The black squares are the mean values inside each bin only for the compilation by Cescutti et al. (2007) and the error bars are the standard deviations. It can be seen that Model C97 reproduces quite well the data, even in the inner part of the disk, since the slope is more pronunced than that of Model C08. As already said, Model C08 is too flat in the inner part of the disk. For this reason we test several parameters in order to better reproduce the inner abundance gradients. In the outer parts of the galactic disk the slope of Model C08 is more pronunced than that of Model C97.

\begin{figure}
\centering
\includegraphics[width=0.45\textwidth]{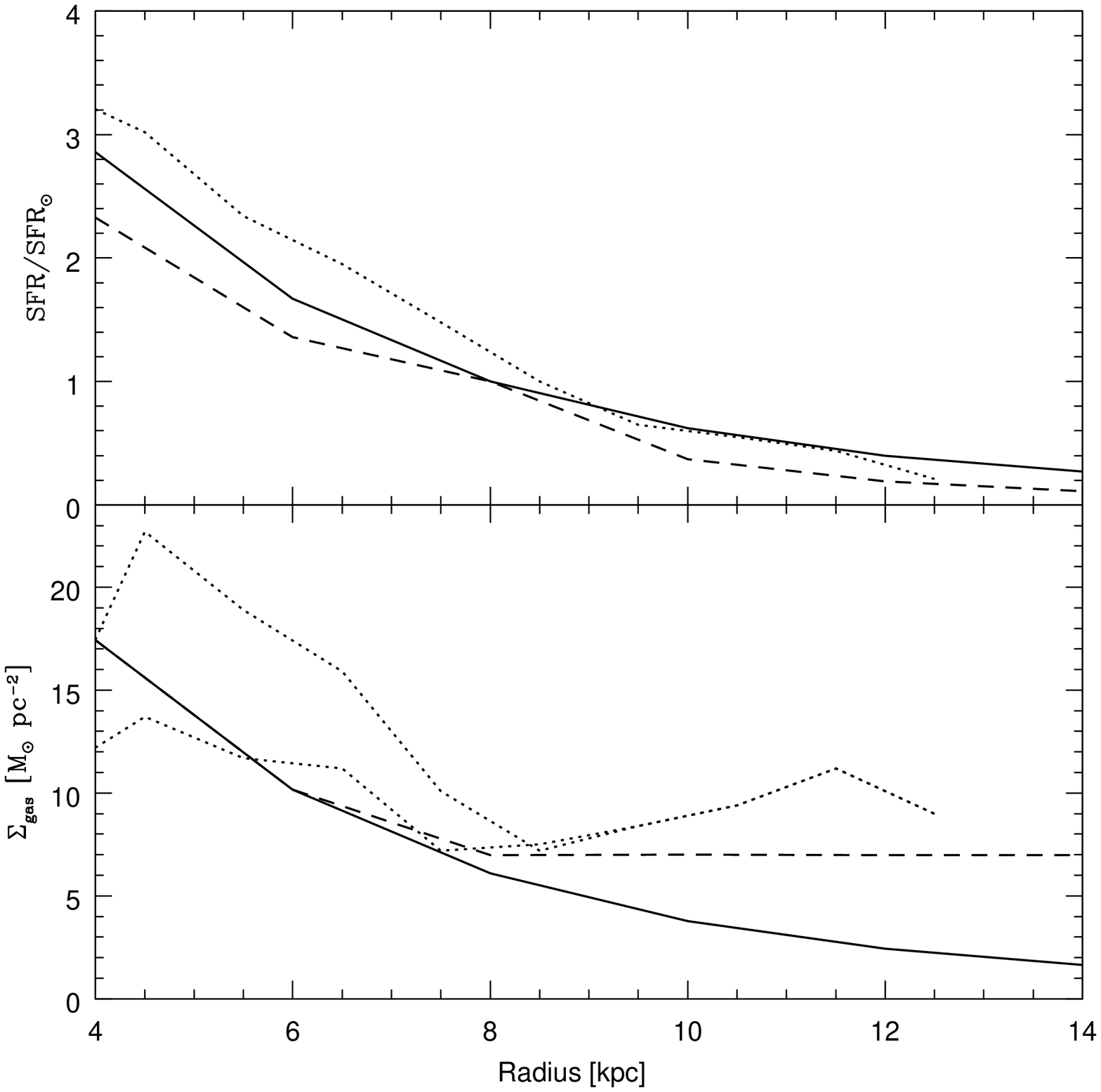}
\caption[]{$SFR/SFR_{\odot}$ and surface gas density vs radius for Model 1 (black solid line) and for Model C08 (red dashed line). The data are the same of Figure \ref{C97.1}.}
\label{cosmo1.1}
\end{figure}

\begin{figure}
\centering
\includegraphics[width=0.45\textwidth]{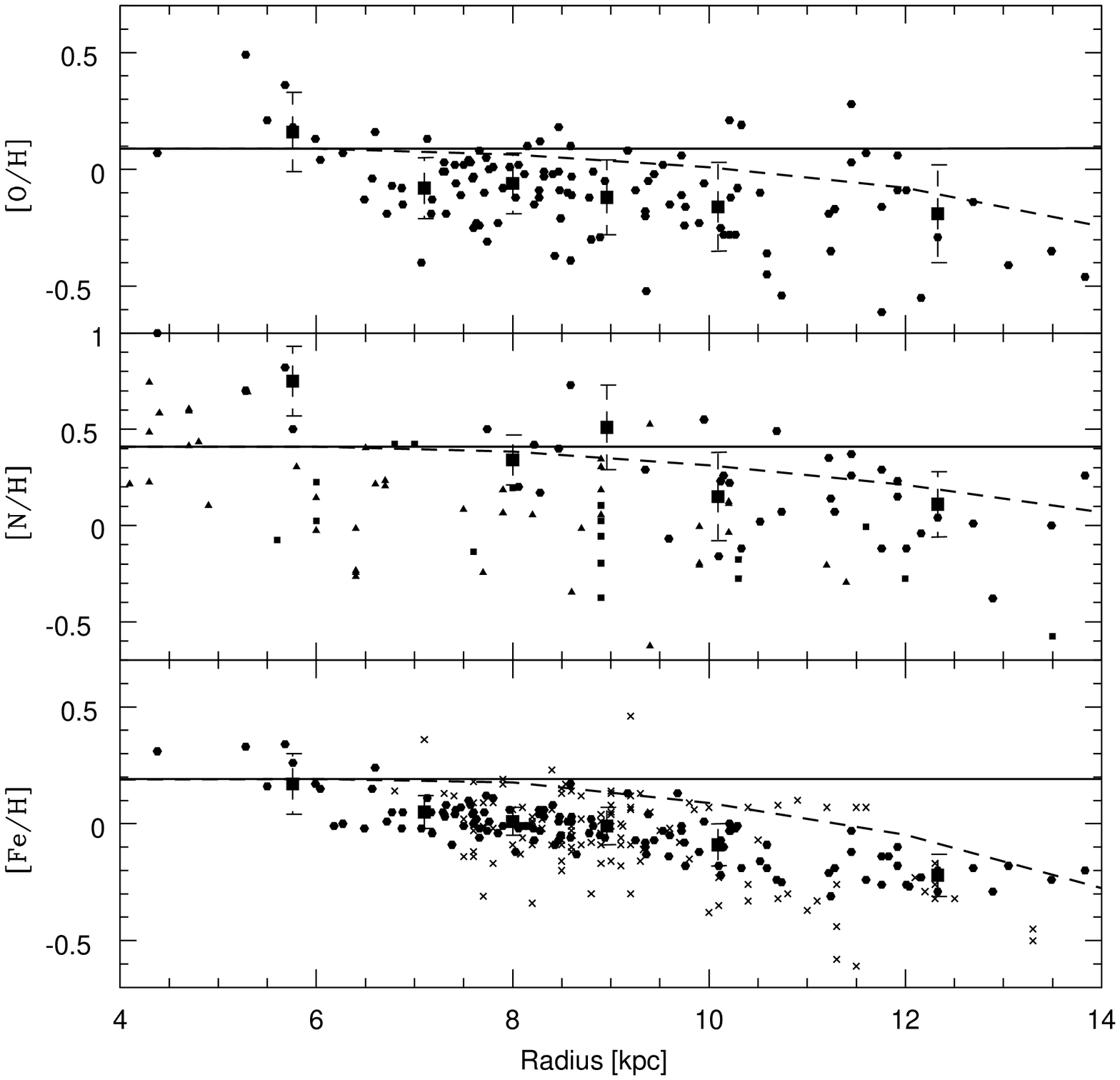}
\caption[]{[O/H], [N/H] and [Fe/H] vs radius for Model 1 (black solid line) and for Model C08 (red dashed line). The data are the same of Figure \ref{C97.2}.}
\label{cosmo1.2}
\end{figure}

Figures \ref{cosmo1.1} and \ref{cosmo1.2} show the results for Model 1, compared with those predicted by Model C08 and with observations. Model 1 is equal to Model C08, except for the lack of the threshold. We can see that Model 1 reproduces very well the $SFR/SFR_{\odot}$, much better than Model C08. However, Model 1 can not reproduce the surface gas density in the outer parts of the disk, since the obtained values are too low respect to the observed values.

Figure \ref{cosmo1.2} shows the abundance gradients for Model 1, compared with the results of Model C08. The data are the same of Figure \ref{C97.2}. In this case the lack of a threshold in the surface gas density makes the abundance gradients completely flat. Therefore, simply removing the threshold from Model C08, it is not possible to reproduce the chemical abundances along the galactic disk, neither in the inner parts nor in the outers. So in the next model (Model 2) we also use a variable star formation efficiency along radius. 

\begin{figure}
\centering
\includegraphics[width=0.45\textwidth]{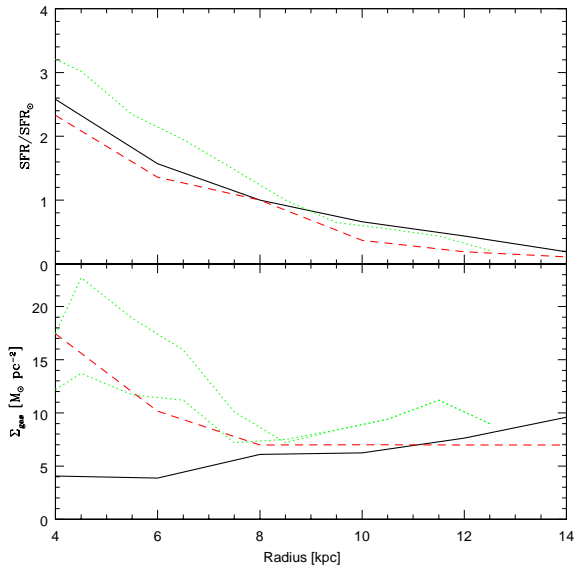}
\caption[]{$SFR/SFR_{\odot}$ and surface gas density vs radius for Model 2 (black solid line) and for Model C08 (red dashed line). The data are the same of Figure \ref{C97.1}.}
\label{cosmo2.1}
\end{figure}

\begin{figure}
\centering
\includegraphics[width=0.45\textwidth]{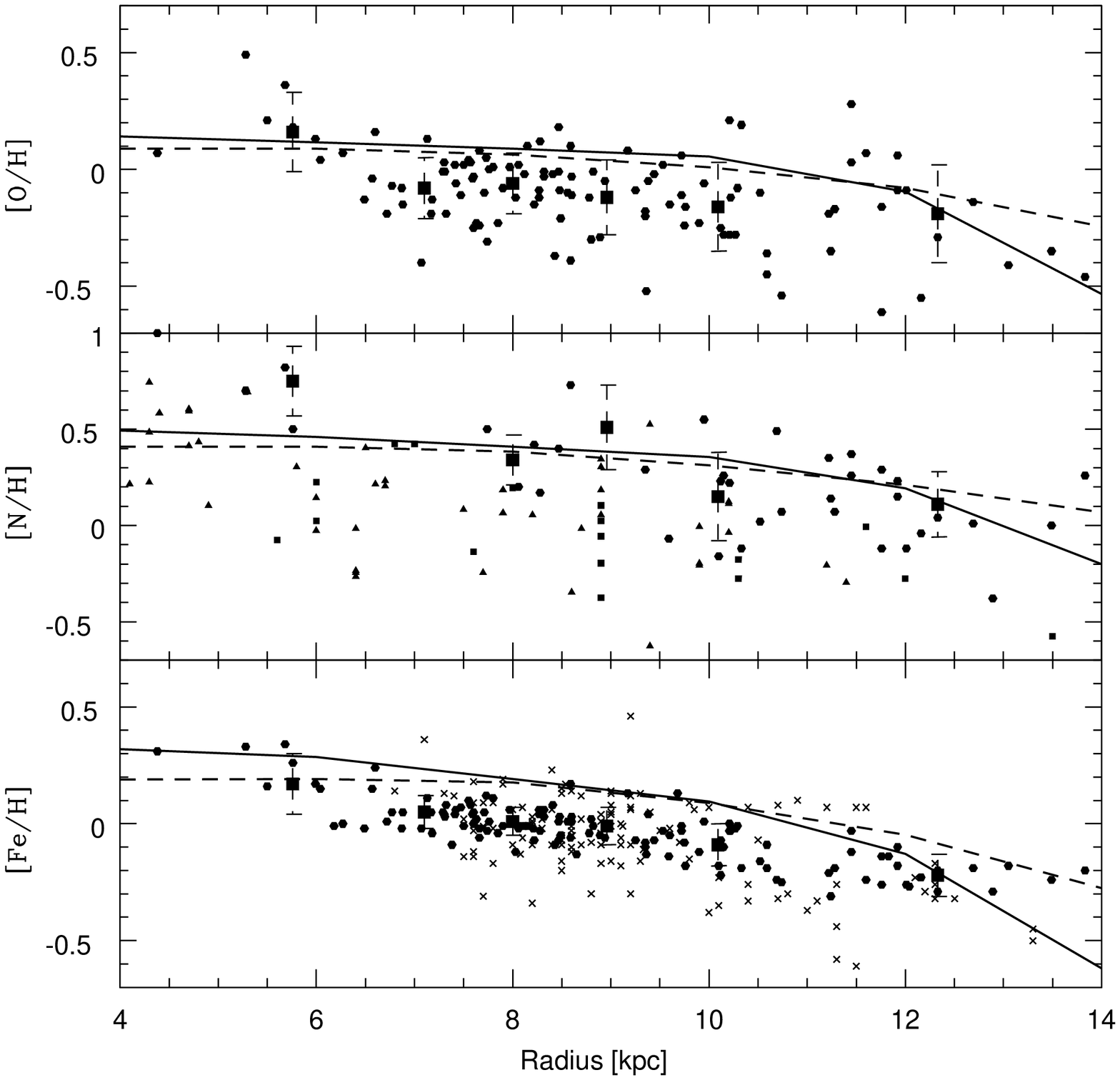}
\caption[]{[O/H], [N/H] and [Fe/H] vs radius for Model 2 (black solid line) and for Model C08 (red dashed line). The data are the same of Figure \ref{C97.2}.}
\label{cosmo2.2}
\end{figure}

In Figures \ref{cosmo2.1} and \ref{cosmo2.2} the results obtained for Model 2, compared with those predicted by Model C08 and with observations, can be seen. Model 2 has the same infall law of Model C08 but in this case there is no threshold during the formation of the galactic disk. Moreover, in Model 2 we use a star formation efficiency which is a function of radius. 

In Figure \ref{cosmo2.1}, upper panel, it can be seen that Model 2 is in better agreement with the data then Model C08, especially in the outer part of the disk. Model C08 has a $SFR/SFR_{\odot}$ equal to zero in the outer parts of the disk, and it is due to the presence of the threshold. From the bottom panel of Figure \ref{cosmo2.1} we can see that Model C08 is in quite good agreement with the data, whereas Model 2 predicts a surface gas density at variance with the observations: it predicts a too low surface gas density in the inner disk because of the high SF efficiency. 

Figure \ref{cosmo2.2} shows the evolution of [O/H], [N/H] and [Fe/H] along the radius for Models C08 and 2. Using no threshold and a star formation efficiency variable with radius again we cannot reproduce, as we hoped, the gradients in the inner parts of the disk. We can only increase the slope of the gradients, relative to Model C08 in the outer parts. 

Since it is not possible to well reproduce at the same time the surface gas density and the abundance gradients along the galactic disk using the cosmologically derived infall law by C08 without the threshold or without the threshold and adopting a SF efficiency variable along radius, we now try a new approch. As we said in section 4, model C08 assumed that the derived infall law has the same functional form for the whole Milky Way, but that the normalization constant is different for different Galactic regions. 

\subsection{Results for the cosmological infall law with inside-out}
We then  tried to derive infall laws having different functional form along the disk. To do this we divided the galactic disk in three zones. The first includes the central 5 kpc of the disk, the second is a radial shell starting at 5 kpc and ending at 10 kpc whereas the third shell has inner and outer radii of 10 and 15 kpc. Having the coordinates of each DM particle of the main progenitor at every redshift, we calculated the total mass included in every shell, deriving three different cosmological infall laws for the three zones. It is well-known (Helmi et al. 2003 and references therein) that the build-up of a Galaxy-sized DM halo proceeds in an inside-out fashion, with the mass of the inner part of the halo being in place at high redshift, while its outer part still accretes mass up to low redshift. We wanted to determine if a cosmological, standard resolution simulation can still capture such a behaviour. This in order to verify if it is possible to use such a simulation to build a cosmological inside-out barionic accretion law, without resorting to high-resolution resimulation of single halos with DM only or with more comples physics. It is also known that re-simulation of single halos, with gas physics, a star formation prescription and some effective form of SN energy feedback can reproduce an inside-out formation of disk galaxies (Governato et al. 2007, Abadi et al. 2003a,b). Such numerical simulations are however computationally very expensive and the properties of the simulated galaxy somehow depend on the sub-grid prescription for the various astrophysical processes. Our aim here is to see whether we can obtain similar results using a much simpler numerical approach.

\begin{figure}
\centering
\includegraphics[width=0.45\textwidth]{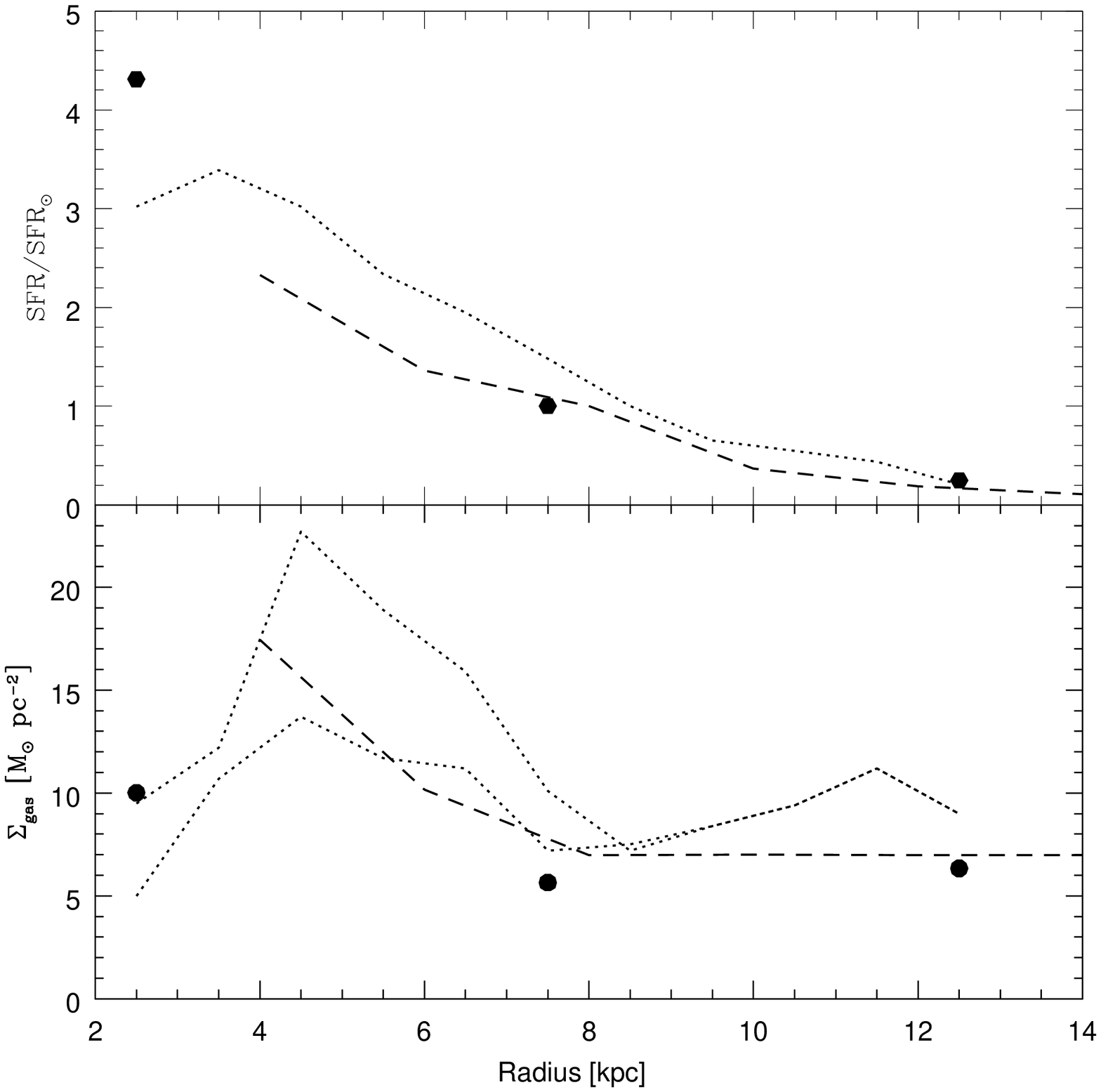}
\caption[]{$SFR/SFR_{\odot}$ and surface gas density vs radius for Model 3 (black points) and for Model C08 (red dashed line). The data are the same of Figure \ref{C97.1}.}
\label{cosmo3.1}
\end{figure}

\begin{figure}
\centering
\includegraphics[width=0.45\textwidth]{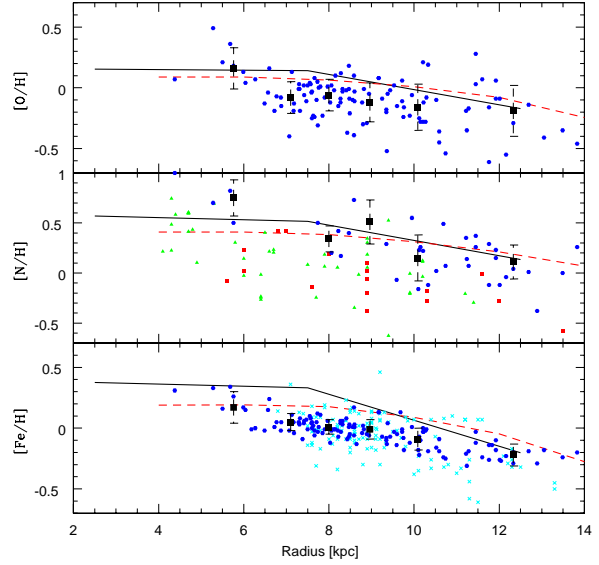}
\caption[]{[O/H], [N/H] and [Fe/H] vs radius for Model 3 (black solid line) and for Model C08 (red dashed line). The data are the same of Figure \ref{C97.2}.}
\label{cosmo3.2}
\end{figure}

Then we tested these laws in the chemical evolution model, using no threshold and a star formation efficiency which changes with radius. In Figures \ref{cosmo3.1} and \ref{cosmo3.2} the results can be seen.

Figure \ref{cosmo3.1} shows that Model 3 reproduces very well the observed SFR, especially between 9 and 12.5 kpc. The slope of the curve of observable data is very similar to that of Model 3. Only at 2.5 kpc the predicted SFR is slightly larger then the observed. However the improvement respect to model C08 is clear and it is mainly due to the lack of a threshold in the superficial gas density. In the bottom panel it can be seen that Model 3 has a too low $\Sigma_{gas}$ respect to the observable data and model C08. However it is interesting to note that the behavior of the curve is quite similar to that of the observable data since the increasing of the superficial gas density observed by Rana (1991) in the outer part of the disk is also predicted by Model 3. Moreover the value at 2.5 kpc is equal to that given by the observable data.

Figure \ref{cosmo3.2} shows the obtained [O/H], [N/H] and [Fe/H] along the radius for Model 3. It can be seen that in the case of O and N the results are quite similar to those of model C08, even if the slope produced by Model 3 is more pronunced. The slope is even more evident in the case of [Fe/H], especially for the outer part of the disk. However, not even  Model 3 can well reproduce the slope in the inner part of the Galaxy, in spite of the fact that the accretion law contains an inside-out effect. This is due to the mild inside-out effect predicted by the infall law which can be attributed to the relatively low resolution that our cosmological simulation can achieve on a single halo, especially at high redshifts. 

\subsection{Changing the parameters in the C97 models}
After this study on the cosmological model by C08, we considered the model by C97, trying to find the best combination of parameters to well reproduce the $SFR/SFR_{\odot}$, the surface gas density and the abundance gradients along the Galactic disk, as done with Model C08. In this case we studied the behavior of the parameters in four different models, comparing every model with Model C08. 

\begin{figure}
\centering
\includegraphics[width=0.45\textwidth]{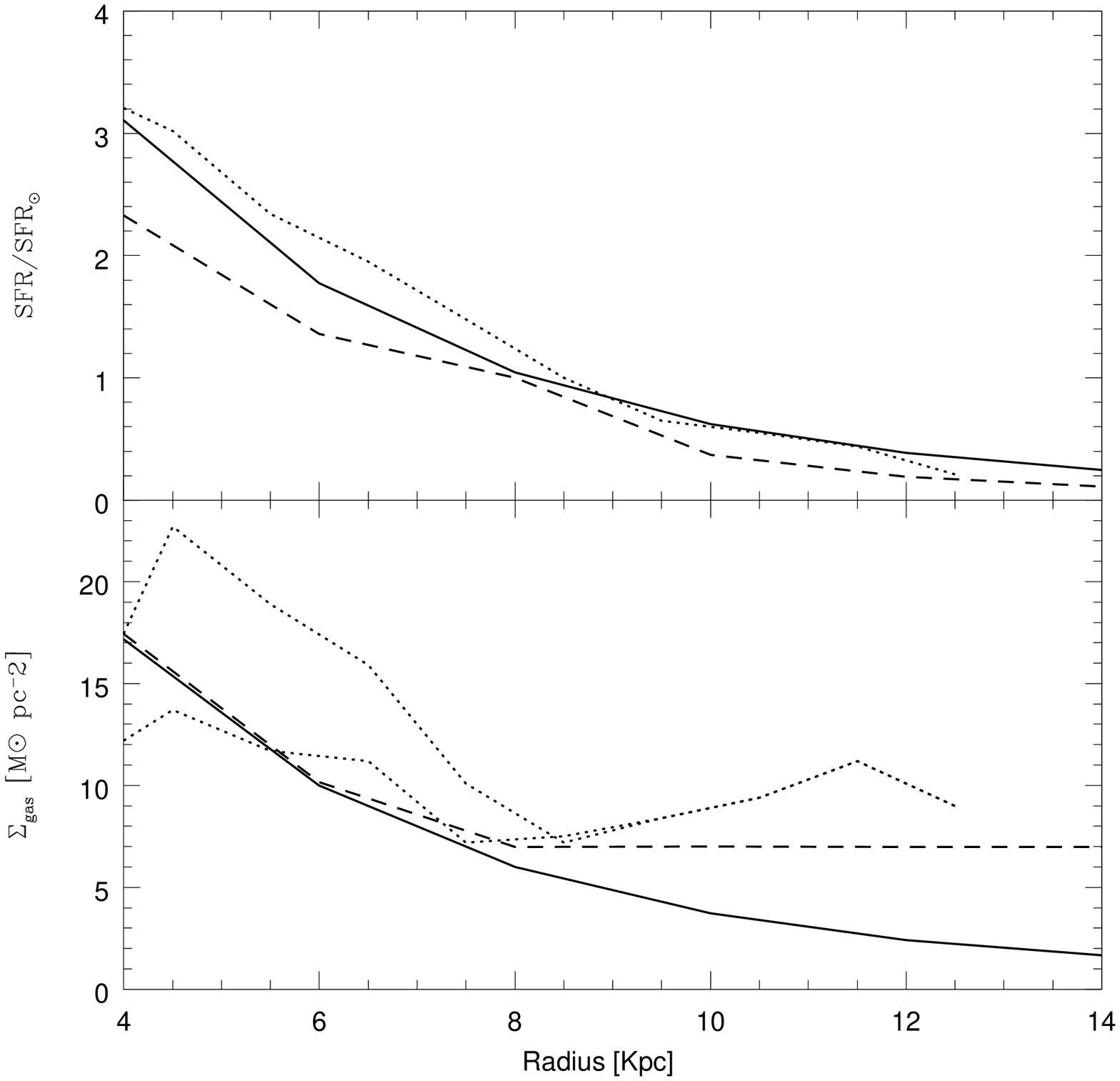}
\caption[]{$SFR/SFR_{\odot}$ and surface gas density vs radius for Model 4 (black solid line) and for Model C08 (red dashed line). The data are the same of Figure \ref{C97.1}.}
\label{4.1}
\end{figure}

\begin{figure}
\centering
\includegraphics[width=0.45\textwidth]{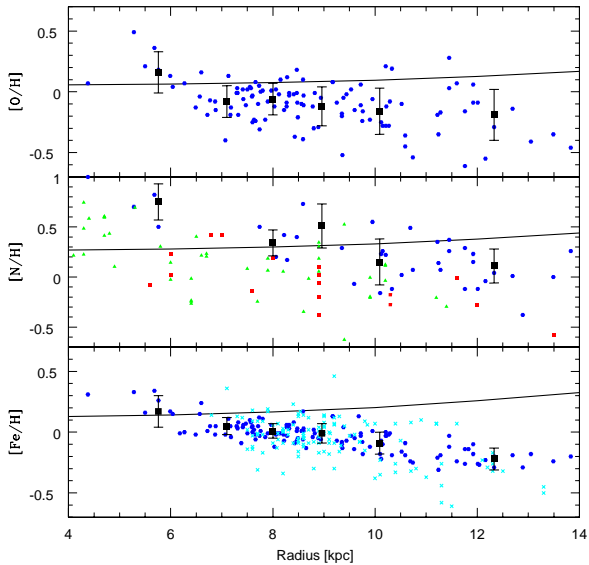}
\caption[]{[O/H], [N/H] and [Fe/H] vs radius for Model 4 (black solid line) and for Model C08 (red dashed line). The data are the same of Figure \ref{C97.2}. It can be seen that in the outer part of the disk the gradients increase. In fact, being the halo surface mass density constant along the galactic disk and since there is no threshold in the surface gas density, the star formation in the halo overcomes that in the outer disk.}
\label{4.2}
\end{figure}

Figures \ref{4.1} and \ref{4.2} show the results for Model 4. This Model has no threshold and the SF efficiency $\nu$, $\tau$ and $\Sigma_{halo}$ are constants. In Figure \ref{4.1} it can be seen that the predicted star formation rate of Model 4 agrees very well with the data by Rana (1991). However $\Sigma_{gas}$ is too low in the outer parts of the disk, even if in the inner parts it agrees quite well with the observed values.      

Figure \ref{4.2} shows the [O/H], [N/H] and [Fe/H] along radius for Model 4. We can see that in the inner disk the abundance gradients are flat, while in the outer parts they even increase, in contrast with the observed data. Therefore, it is not possible to reproduce the abundance gradients without the threshold and when $\nu$, $\tau$ and $\Sigma_{halo}$ are constants. This is due to the fact that with a constant $\tau$ the disk evolution is too quick to allow the formation of abundance gradients, that the star formation never stops (absence of the threshold) and that the halo surface mass density is constant with galactocentric distance. In this case, in fact, the star formation in the halo overcomes that in the outer disk and produces higher abundances than in the more realistic case of a fading surface halo mass density. 

\begin{figure}
\centering
\includegraphics[width=0.45\textwidth]{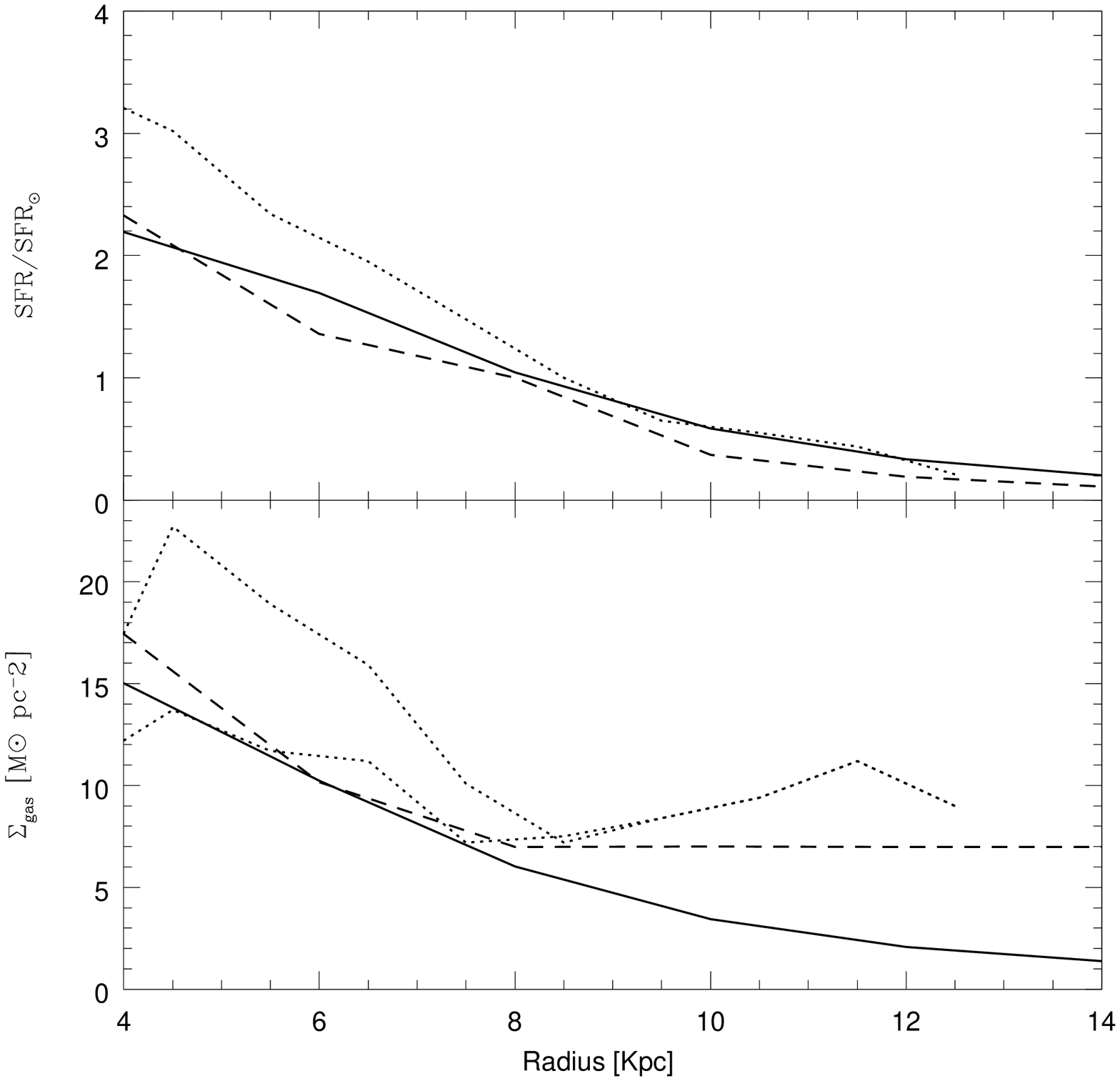}
\caption[]{$SFR/SFR_{\odot}$ and surface gas density vs radius for Model 5 (black solid line) and for Model C08 (red dashed line). The data are the same of Figure \ref{C97.1}.}
\label{5.1}
\end{figure}

\begin{figure}
\centering
\includegraphics[width=0.45\textwidth]{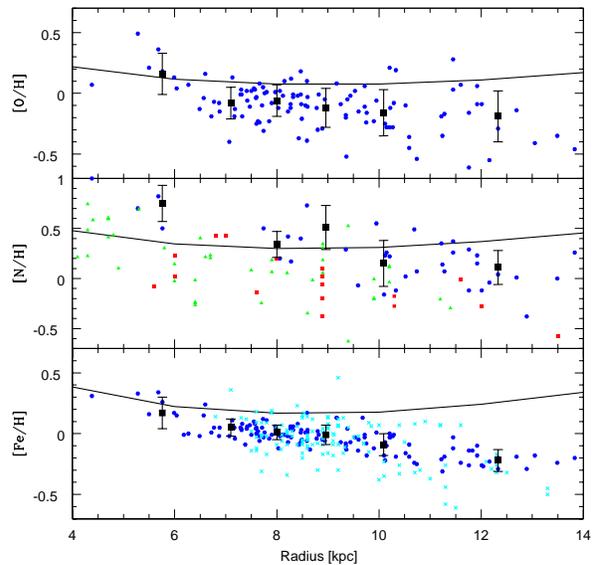}
\caption[]{[O/H], [N/H] and [Fe/H] vs radius for Model 5 (black solid line) and for Model C08 (red dashed line). The data are the same of Figure \ref{C97.2}. This model shows that the presence of a variable $\tau$ along the galactic disk steepens the inner gradients, without however reducing the positive slope in the outer disk, as seen in Fig. \ref{4.2}.}
\label{5.2}
\end{figure}

Figures \ref{5.1} and \ref{5.2} show Model 5, compared to Model C08. Model 5 has no threshold (neither in the halo nor in the disk), a constant star formation efficiency, a variable $\tau$ along the disk and a constant surface halo density. From Figure \ref{5.1} it can be seen that the star formation rate predicted by Model 5 is in quite good agreement with the data by Rana (1991), especially for the outer parts of the galaxy. However the amount of gas is too low, even if compared with the lower values by Rana (1991).

Figure \ref{5.2} shows the abundance gradients along the galactic disk for Model 5. In the inner part of the disk the [O/H], [N/H] and [Fe/H] ratios are not flat as in Model C08, reproducing better the observed data. Nevertheless, beyond 10 kpc the gradients become positive at variance with the observations. So we can say that a model without threshold and with a variable $\tau$ cannot well reproduce the data, especially in the outer disk.

\begin{figure}
\centering
\includegraphics[width=0.45\textwidth]{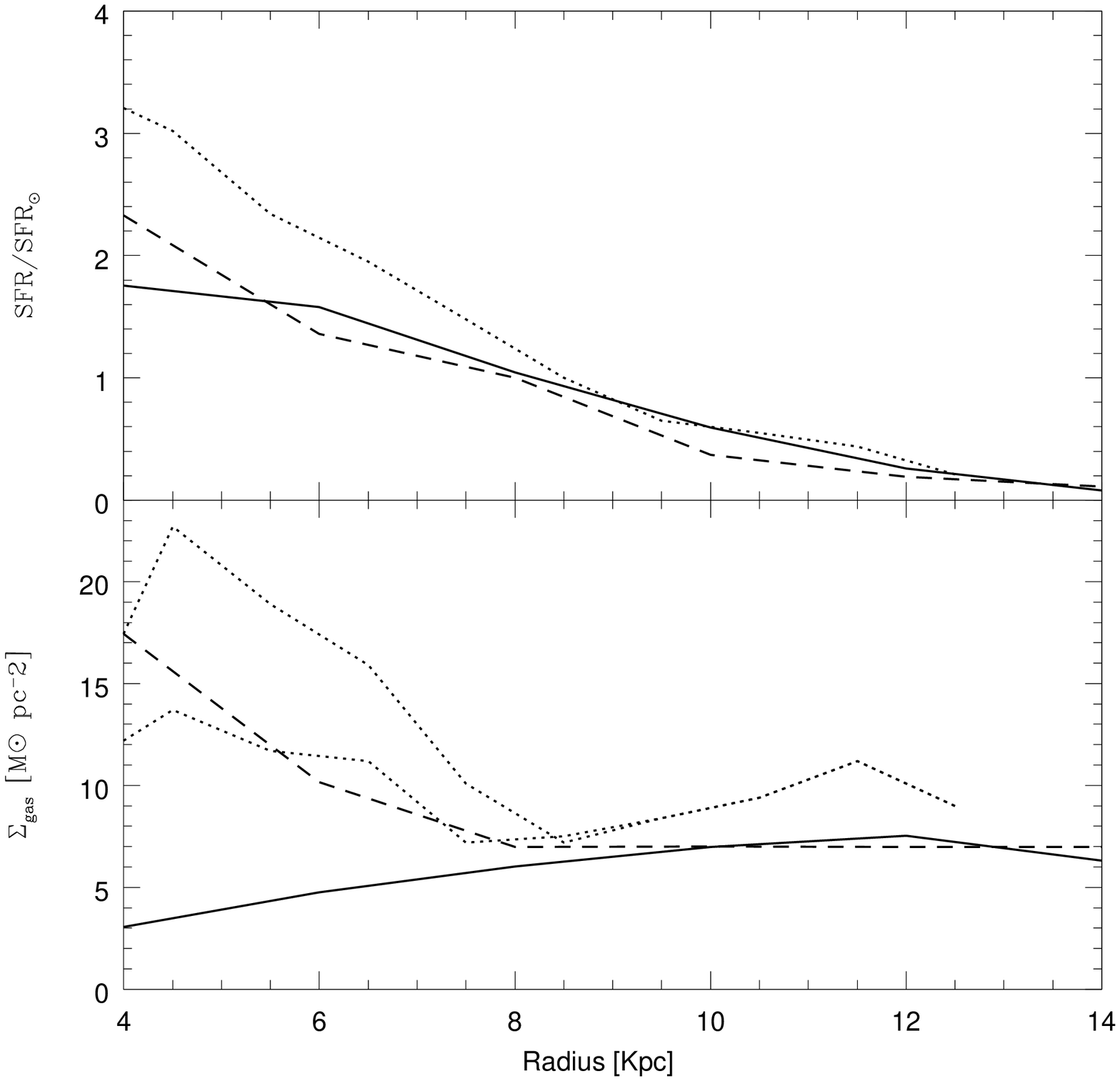}
\caption[]{$SFR/SFR_{\odot}$ and surface gas density vs radius for Model 6 (black solid line) and for Model C08 (red dashed line). The data are the same of Figure \ref{C97.1}.}
\label{6.1}
\end{figure}

\begin{figure}
\centering
\includegraphics[width=0.45\textwidth]{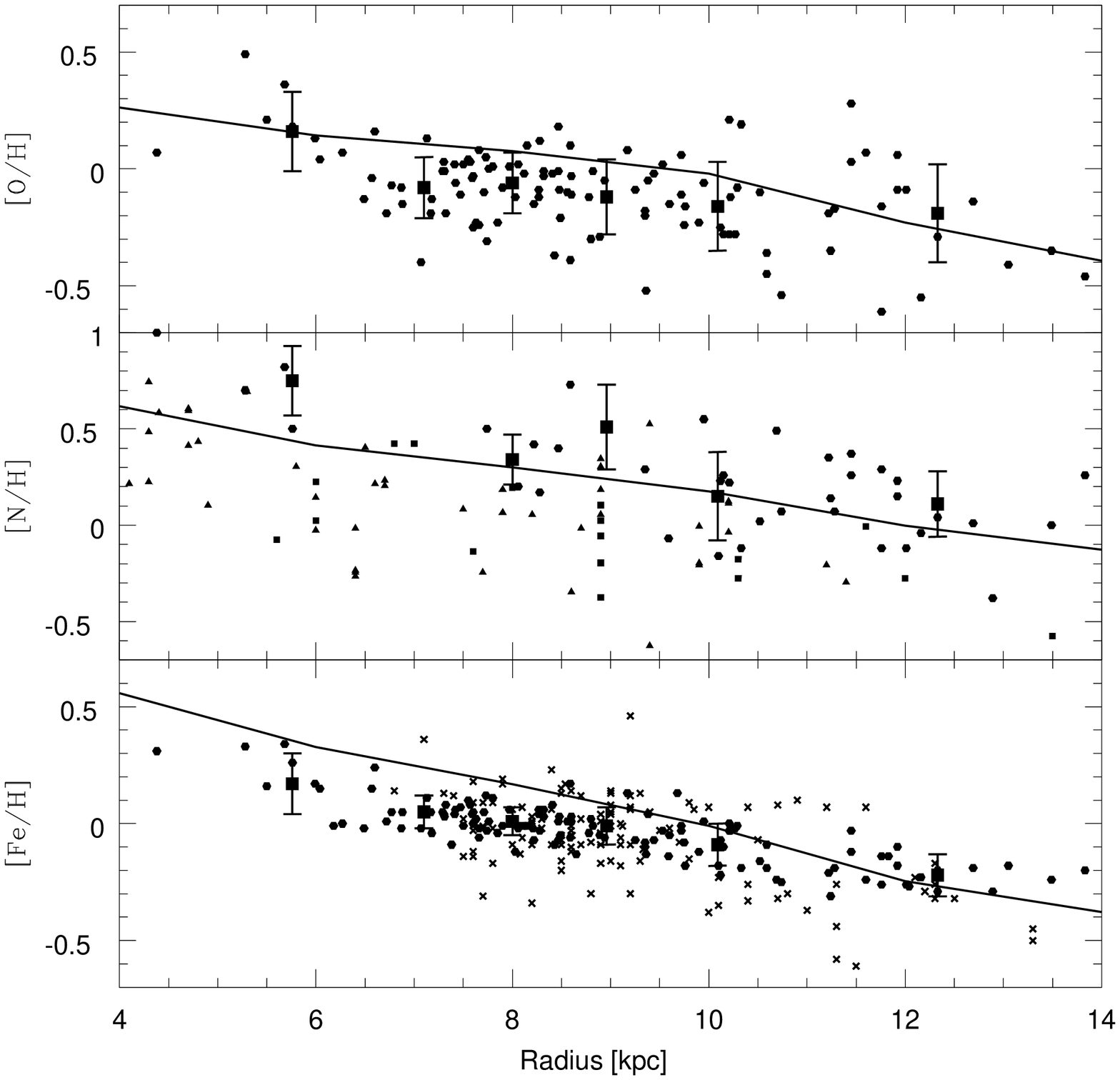}
\caption[]{[O/H], [N/H] and [Fe/H] vs radius for Model 6 (black solid line) and for Model C08 (red dashed line). The data are the same of Figure \ref{C97.2}.}
\label{6.2}
\end{figure}

Figures \ref{6.1} and \ref{6.2} show the $SFR/SFR_{\odot}$, $\Sigma_{gas}$ and the [O/H], [N/H] and [Fe/H] gradients for Model 6. This model has no threshold, a star formation efficiency which changes strongly with radius, an inside-out prescription and a surface halo density which is a function of radius. From Figure \ref{6.1} it can be seen that the SFR is in good agreement with the data, especially in the outer parts of the disk. On the other hand, the surface gas density increases along radius, at variance with the observations which show that there is more gas in the inner parts of the disk.  

In Figure \ref{6.2} we can see that the abundance gradients have a slope which is quite similar to the slope of the data, mainly for the [Fe/H] ratio. Even in the inner parts the gradients are in quite good agreement with the data. However, we cannot say that this model can reproduce all the observations, since it cannot reproduce the correct behavior of $\Sigma_{gas}$.

\begin{figure}
\centering
\includegraphics[width=0.45\textwidth]{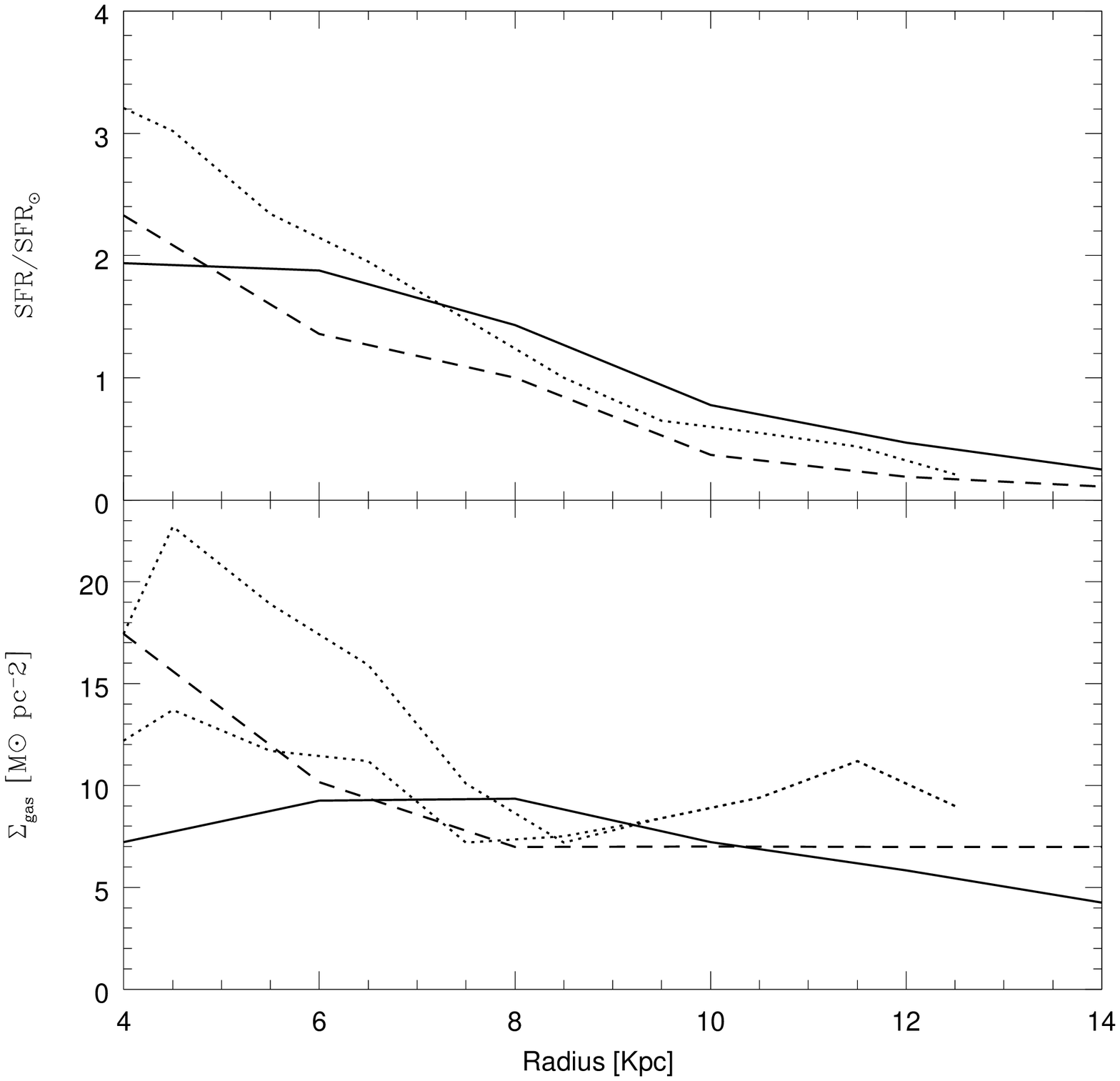}
\caption[]{$SFR/SFR_{\odot}$ and surface gas density vs radius for Model 7 (black solid line) and for Model C08 (red dashed line). The data are the same of Figure \ref{C97.1}.}
\label{7.1}
\end{figure}

\begin{figure}
\centering
\includegraphics[width=0.45\textwidth]{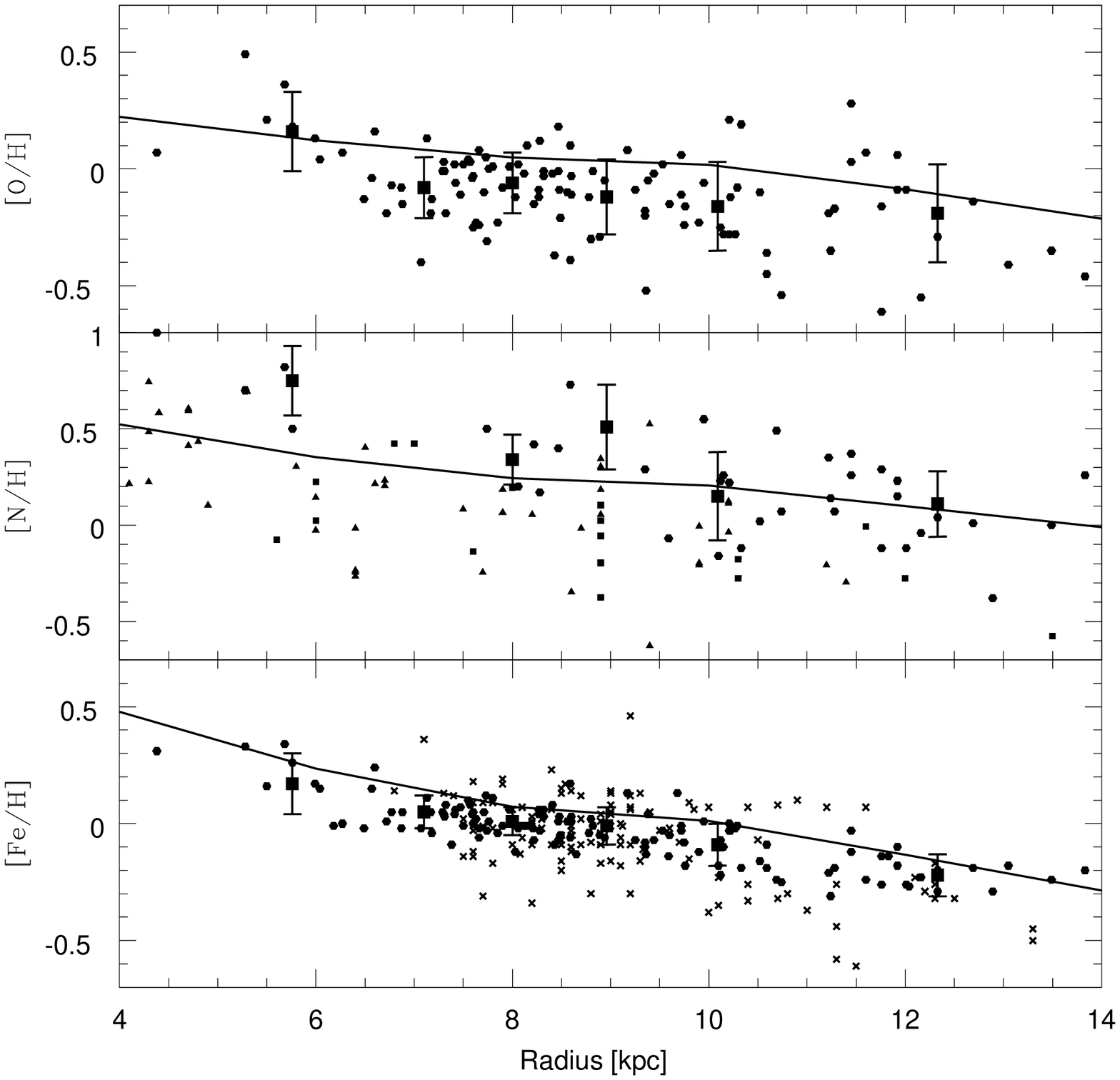}
\caption[]{[O/H], [N/H] and [Fe/H] vs radius for Model 7 (black solid line) and for Model C08 (red dashed line). The data are the same of Figure \ref{C97.2}.}
\label{7.2}
\end{figure}

The last model we present is Model 7. Its results are shown in Figures \ref{7.1} and \ref{7.2}. This model has no threshold, a star formation efficiency which follows that one given by Boissier \& Prantzos (1999) and no inside-out prescription. In this model we assume only one episode of accretion where halo forms first and then the disk follows. It differs from all the other models also because it adopts a simple Schmidt law for star formation, like in eq. (2). It can be seen that the SFR agrees well with the data by Rana (1991), while the surface gas density is too low in the inner part of the Galactic disk. This is again due to a too high SF efficiency in the inner disk.

From Figure \ref{7.2} we can see that Model 7 predicts abundance gradients which are in good agreement with the observed data, especially for the outer parts of the Galactic disk. Therefore we can say that this is the only case with no threshold and no inside-out prescription where it is possible to well reproduce abundance gradients, although the surface gas density is too low in the inner part of the Galaxy.  

\section{Conclusions}

Starting from the model by Colavitti et al. (2008), adopting a cosmologically motivated infall law resembling the one of C97 for the solar neighbourhood but not predicting an inside-out formation of the Galactic disk, and in which the authors were not able to reproduce the [O/H] gradient in the inner parts of the disk, and from the model by C97, we tested several models in order to find the best values of four parameters to best reproduce the $SFR/SFR_{\odot}$, $\Sigma_{gas}$ and the [O/H], [N/H] and [Fe/H] gradients along the Galactic disk. These four parameters are the threshold gas density for the star formation, the star formation efficiency $\nu$, the inside-out prescription $\tau(r)$ (i.e. the variation of the timescale for the formation of the Galactic disk) and the surface halo mass density. 

In table 2 we summarize model successes and failures for abundance trends, surface gas densities and star formation rates.

\begin{table*}
\caption{Models successes and failures. The second column shows the successes while the third column the failures.}
\centering
\begin{tabular}{c|c|c}
\noalign{\smallskip}
\hline
\hline
\noalign{\smallskip}
Model & Successes & Failures \\
\noalign{\smallskip}
\hline
\noalign{\smallskip}
1 & good agreement for the $SFR/SFR_{\odot}$ & $\Sigma_{gas}$ not reproduced in the outer Galactic disk \\
  & & abundance gradients completely flat \\
\noalign{\smallskip}
\hline
\noalign{\smallskip}
2 & good agreement for the $SFR/SFR_{\odot}$ & too low $\Sigma_{gas}$ in the inner disk \\
  & increased slope of the gradients in the outer disk & abundance gradients too flat in the inner disk \\
\noalign{\smallskip}
\hline
\noalign{\smallskip}
3 & good agreement for the $SFR/SFR_{\odot}$ & too low $\Sigma_{gas}$ \\
  & $\Sigma_{gas}$ increases in the outer part of the disk & abundance gradients too flat in the inner disk \\
  & slope of abundance gradients more pronunced in the outer disk & \\ 
\noalign{\smallskip}
\hline
\noalign{\smallskip}
4 & good agreement for the $SFR/SFR_{\odot}$ & $\Sigma_{gas}$ not reproduced in the outer Galactic disk \\
  & & abundance gradients increase along the galactic disk \\
\noalign{\smallskip}
\hline
\noalign{\smallskip}
5 & good agreement for the $SFR/SFR_{\odot}$ & $\Sigma_{gas}$ not reproduced in the outer Galactic disk \\
  & slope of abundance gradients more pronunced in the inner disk & abundance gradients increase in the outer Galactic disk \\
\noalign{\smallskip}
\hline
\noalign{\smallskip}
6 & good agreement for the $SFR/SFR_{\odot}$ & $\Sigma_{gas}$ increases along radius \\
  & abundance gradients in good agreement with the observed data & \\
\noalign{\smallskip}
\hline
\noalign{\smallskip}
7 & good agreement for the $SFR/SFR_{\odot}$ & $\Sigma_{gas}$ too low in the inner part of the Galactic disk \\
  & abundance gradients in good agreement with the observed data & \\ 
\noalign{\smallskip}
\hline
\hline
\end{tabular}
\end{table*}  

Our conclusions can be summarized as follows:

\begin{itemize}

\item We found that it is impossible to fit all the disk constraints at the same time without assuming an inside-out formation for the Galactic disk together with a threshold in the gas density for the star formation rate. In particular, the inside-out formation is important to reproduce the right slope of the abundance gradients in the inner disk, whereas the threshold gas density is important to reproduce the slope of the gradients in the outer disk. Models with a constant timescale for the disk formation (no inside-out) cannot reproduce the slope in the inner disk. On the other hand, models with no inside-out mechanism can well reproduce the distributions of the star formation rate and the gas density along the disk. A good way of testing the inside-out disk formation is to test whether there is a size and luminosity evolution of disks at high redshift. Roche et al. (1998), by studying galaxies in the reshift range 2.5-3.0, concluded that the larger increase of the surface brightness relative to the luminosity at high redshift is an indication of an inside-out disk formation. They tested their results by means of the Chiappini et al. (1997) model. Another important issue concerns the existence of a threshold gas density for star formation. Recently, Boissier et al. (2006), by means of Galex data, suggested that the threshold might not exist or be much lower that thought before. Note, however, that the low level of star formation observed with GALEX at the outer edges of disks does not produce OB associations, hence no massive stars which are main responsible for enriching the ISM. Therefore, the abundance gradients really measure the threshold for massive star formation.

\item In the framework of models with no inside-out mechanism such as the model proposed by Colavitti et al. (2008) we tested the effect of the efficiency of star formation varying with the galactocentric radius, being higher in the innermost than in the outermost regions of the Galactic disk. This assumption, even without the threshold in the surface gas density, can produce gradients with the right slope both in the inner and outer regions of the Galactic disk but it fails in reproducing the gas density distribution along the disk.

\item A cosmological model without threshold, with a star formation efficiency changing with radius and with a very simple inside-out prescription, can well reproduce the SFR along the disk and the behavior of the surface gas density. Moreover it can well reproduce the outer abundance gradients. In the inner part of the disk it produces too flat gradients, even if the slope is more pronunced then that of Model C08.

\item Therefore, we conclude that to reproduce at the same time the abundance, star formation rate and surface gas density gradients along the Galactic disk it is necessary to assume an inside-out formation for the disk. The threshold in the gas density is not necessary and the same effect could be reached by assuming a variable efficiency of star formation. The inclusion of radial flows without an inside-out formation of the disk would have the same effect as the theshold gas density without inside-out effect (Portinari \& Chiosi, 2000).

\item It is clear from our numerical simulations that while the process of formation of the Galactic disk has a major impact on the formation of abundance gradients, the specific adopted SF law has a major effect on the gas distribution along the disk. This effect could be used to infer the SF law from the gas distributions along disks of spirals.

\item As far as the Colavitti et al. (2008) model is concerned, we conclude that our cosmological simulation has not a sufficient resolution to be able to capture the inside-out formation of a galaxy-sized DM halo. High resolution cosmological simulations and/or resimulation of a single DM halo should be performed in order to understand if a suitable baryon accretion law can be inferred from collisionless simulations, or if the full gas physics and sub-grid treatment of astrophysical processes as star formation and energy feedback from SNe is unavoidably needed to get a realistic description of the formation of a galaxy.

\end{itemize}

\section{Acknowledgments}

We thank Francesco Calura for some helpful suggestions. E. C. and F. M. aknowledge funds from MIUR, COFIN 2007, prot. N. 2007JJC53X. G. C. aknowledges financial support from the Fondazione Cassa di Risparmio di Trieste. We also thank the unknown referee for the suggestions.

\label{lastpage}

\end{document}